\documentclass[twocolumn]{aastex631}
\usepackage{graphicx}	
\usepackage{amsmath}	
\usepackage[T1]{fontenc}
\usepackage{newtxtext,newtxmath}

\usepackage{amssymb}	
\usepackage{gensymb}
\usepackage{color}
\usepackage{natbib}
\usepackage{enumitem}
\usepackage{mathtools}
\usepackage{threeparttable}
\usepackage{textcomp,booktabs}   
\usepackage{multirow}
\usepackage{verbatim}

\usepackage[mathlines]{lineno} 

\begin{document}

\title{Transitional Dynamics: Unveiling the Coexistence and Interplay of Type-B and Type-C QPOs in MAXI J1348-630}

\correspondingauthor{Zhen Yan \& Ren-Yi Ma}
\email{zyan@shao.ac.cn, ryma@xmu.edu.cn}
\author{Xin-Lei Wang}
\affil{Department of Astronomy and Institute of Theoretical Physics and Astrophysics, Xiamen University, Xiamen, Fujian 361005, China}
\affil{SHAO-XMU Joint Center for Astrophysics,  Xiamen, Fujian 361005, China}

\author{Zhen Yan}
\affil{Shanghai Astronomical Observatory, Chinese Academy of Sciences, 80 Nandan Road, Shanghai 200030, China}
\affil{SHAO-XMU Joint Center for Astrophysics,  Xiamen, Fujian 361005, China}

\author{Fu-Guo Xie}
\affil{Shanghai Astronomical Observatory, Chinese Academy of Sciences, 80 Nandan Road, Shanghai 200030, China}
\affil{SHAO-XMU Joint Center for Astrophysics,  Xiamen, Fujian 361005, China}

\author{Jun-Feng Wang}
\affil{Department of Astronomy and Institute of Theoretical Physics and Astrophysics, Xiamen University, Xiamen, Fujian 361005, China}
\affil{SHAO-XMU Joint Center for Astrophysics,  Xiamen, Fujian 361005, China}

\author{Ya-Xing Li}
\affil{Shanghai Astronomical Observatory, Chinese Academy of Sciences, 80 Nandan Road, Shanghai 200030, China}
\affil{University of Chinese Academy of Sciences, 19A Yuquan Road, Beijing 100049, China}

\author{Ren-Yi Ma}
\affil{Department of Astronomy and Institute of Theoretical Physics and Astrophysics, Xiamen University, Xiamen, Fujian 361005, China}
\affil{SHAO-XMU Joint Center for Astrophysics,  Xiamen, Fujian 361005, China}

\begin{abstract}

Based on broadband timing analysis of \textit{Insight}-HXMT and \textit{NICER} data from the 2019 outburst of the black hole X-ray binary (BHXRB) MAXI J1348-630, we report the detection of the coexistence and competitive interplay between type-C and type-B quasi-periodic oscillations (QPOs). Specifically, the two QPO types were detected simultaneously but exhibited distinct energy dependencies: the type-C QPO was dominant in the hard X-ray band (10–30 keV), while the type-B QPO was more prominent in the soft X-ray band (1–10 keV). Further analysis reveals that the emergence of the type-C QPO suppresses the strength of the type-B QPO, particularly above 10 keV. Tracking the detailed evolution of these QPOs, we find that the weakening of the type-B QPO coincides with the strengthening of the type-C QPO, indicating a competitive interaction between them. These findings reveal a more complex relationship between type-B and type-C QPOs than previously recognized, suggesting they originate from distinct but interacting physical components within the accretion flow and/or jet, and providing new insights into the dynamics of accretion state transitions.

\end{abstract}



\section{Introduction}
\label{sec:intro}

The luminosity and X-ray spectrum of black hole X-ray binaries (BHXRBs) typically undergo distinct stages during an outburst. The outburst starts from the hard state (HS), moves to the hard-intermediate state (HIMS) and soft-intermediate state (SIMS), then transitions to the soft state (SS), and finally returns to the HS following quenching \citep{Mendez97, Homan_2005, Remillard_2006, Belloni10, Belloni11, Plant14, Belloni_2016}.

Beyond this long-term evolution, BHXRBs also exhibit short-term variability. This variability can be effectively investigated using the Fourier transform, which converts the X-ray light curve from the time domain to the frequency domain and generates the power density spectrum (PDS) \citep{vanderKlis1988}. 
The PDS displays the intensity of variability across different frequencies; sharp peaks occasionally emerge, which correspond to quasi-periodic oscillations (QPOs) \citep[e.g.,][]{Nowak00, Casella04, Belloni_2014}.

QPOs are characterized by three parameters: the central frequency, the quality factor \( Q \) (defined as the ratio of the central frequency to the full width at half maximum, FWHM), and the fractional rms (root mean square) amplitude. As a normalized measure of variability intensity, the fractional rms of QPO quantifies the relative intensity of QPO with respect to the total mean intensity \citep{VanDerKlis89,Ingram19}. Based on their central frequencies, QPOs are generally classified into two categories: low-frequency QPOs (LFQPOs), with frequencies ranging from a few mHz to tens of Hz \citep{psaltis_1999,Nowak00,Belloni02a, Casella04,Belloni05}; and high-frequency QPOs (HFQPOs), with frequencies spanning tens to hundreds of Hz \citep{Morgan97,Remillard99a,Belloni01,Homan01, Remillard02, Altamirano12HF}. 

LFQPOs are typically classified into three types: A, B, and C \citep{Casella04, Casella05, Motta_2011}. Type-A QPOs are rarely detected in the SIMS and SS, characterized by a weak, broad peak in the PDS \citep[e.g.][]{zhang_2023}. 
Type-B QPOs are detected in the SIMS at a typical frequency range of 4--6 Hz, usually accompanied by weak red noise (few percent fractional rms or less) \citep[e.g.,][]{Remillard02,Stevens_2016,Belloni_2016,Gao_2017}.
Type-C QPOs are the most common LFQPOs in BHXRBs, with strong fractional rms amplitudes (3--16\%) and high \( Q \) values \citep{Belloni10,Ingram19}. The frequency of type-C QPOs may increase from tens of mHz to around 10 Hz \citep[e.g., ][]{Casella04,Belloni05,Motta_2011, Huang_2018,Buisson2019}. Although type-C QPOs can appear in all states, they are most frequently detected in the HS and HIMS, associated with strong band-limited noise (BLN) \citep[e.g.,][]{Belloni02a}. Thus the disappearance of the type-C QPO and the appearance of the type-B QPO is usually identified as the signature of HIMS-to-SIMS transition \citep{Belloni10,Belloni_2016}. 
For decades, many studies have focused on different types of QPOs in BHXRBs because they serve as key diagnostics of the structure and dynamics of the inner accretion flow and jet around the BH \citep[see reviews in ][]{Belloni_2014,Ingram19}.

MAXI J1348-630 was discovered during its 2019 outburst \citep{Yatabe2019}.
From January 2019, MAXI J1348-630 underwent a complete outburst that lasted approximately 110 days, followed by several mini-outbursts. 
Based on spectral-timing analyses of data from the Neutron Star Interior Composition Explorer (\textit{NICER}), \citet{Zhang2020} suggested the source as a black hole system and detected all three typical LFQPOs.
Recent studies on QPOs in MAXI J1348-630 suggest a complex interplay between accretion flow dynamics, coronal geometry, and jet activity. Observations indicate that type-B QPOs are associated with jet phenomena, supported by bidirectional phase lags (i.e., both higher- and lower-energy photons lag behind the reference band 2--2.5 keV) explained by large-scale jet structures \citep{Belloni2020} and the emergence of an additional Comptonization component linked to the jet base \citep{Zhang2021}. Rapid transitions between type-C and type-B QPOs further imply that type-B QPOs may stem from the precession of a weak, small-scale jet, while type-C QPOs likely originate from an extended corona or a compact, steady jet \citep{Liu_2022}. On the other hand, a Comptonization modeling of both type-B and type-C QPOs suggests the corona's geometry evolves across different spectral states, transitioning from a horizontally extended structure in the HS to a vertically extended one during the SIMS \citep{Garc2021,Alabarta24}.

In this paper, we perform a detailed timing analysis of MAXI J1348-630
across a broad X-ray band (1--200 keV) using combined data from \textit{Insight}-HXMT and \textit{NICER}. We report the discovery of the coexistence and interplay between type-C and type-B QPOs. In the following sections, we describe the observations and data reduction in \autoref{sec:data}, present the evidence and details of our discovery in \autoref{sec:results}, discuss the implications of this discovery in \autoref{sec:discussion}, and conclude our work in \autoref{sec:conclusion}.


\section{OBSERVATION AND DATA REDUCTION}
\label{sec:data}
\textit{Insight}-HXMT is an X-ray astronomical satellite with a wide energy coverage from 1 to 250 keV. 
It is equipped with three main payloads: the Low Energy X-ray Telescope (LE); the Medium Energy X-ray Telescope (ME); and the High Energy X-ray Telescope (HE) \citep{zhang_overview_2020}.
Each observation from \textit{Insight}-HXMT contains several data segments referred to as "exposures". 
\textit{NICER} is a soft X-ray telescope mounted on the International Space Station (ISS), launched in June 2017. \textit{NICER} operates in the 0.2--12 keV energy band, providing a high effective area and timing precision of approximately 100 ns.

For this work, we primarily analyzed data from three payloads of \textit{Insight}-HXMT. Given the high data quality of \textit{NICER} in the soft X-ray ($<$ 10 keV), we also utilized \textit{NICER} data as a supplement. Hereafter, "LE" (low energy), "ME" (medium energy), and "HE" (high energy) specifically denote the energy bands of the three \textit{Insight}-HXMT payloads: 1--10 keV for LE, 10--30 keV for ME, and 27--200 keV for HE.

\begin{figure*}
\centering
\includegraphics[width=0.9\linewidth]{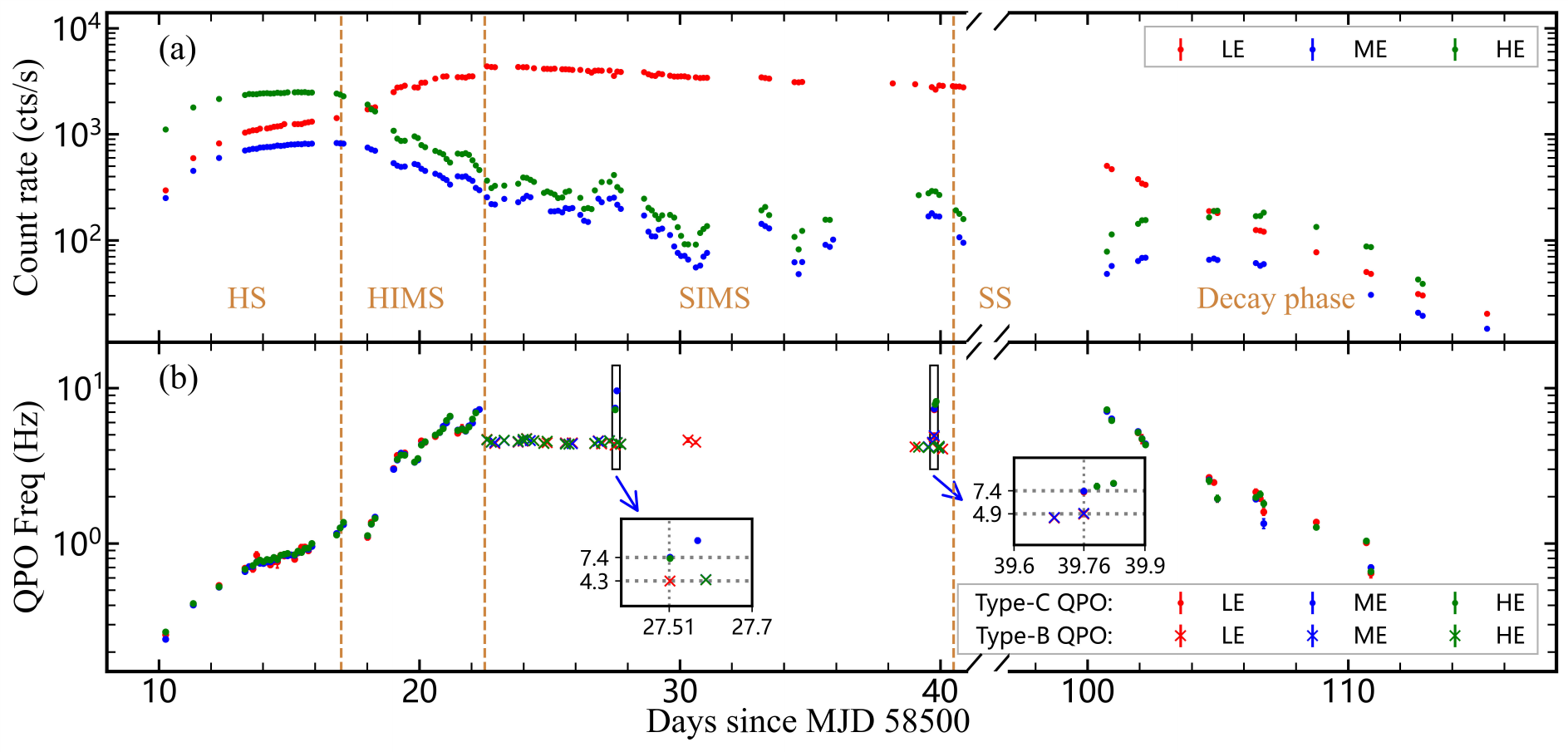}
\caption{An overview of the light curves and QPO frequency evolution during the 2019 outburst of MAXI J1348-630 using \textit{Insight}-HXMT data. In panel (b), 
the inset panels zoom in on the details around MJD 58527.5 and 58539.8, where type-C and type-B QPOs were simultaneously detected.}
\label{fig:QPO-evolution}
\end{figure*}

\begin{figure*}
\centering
\includegraphics[width=1\linewidth]{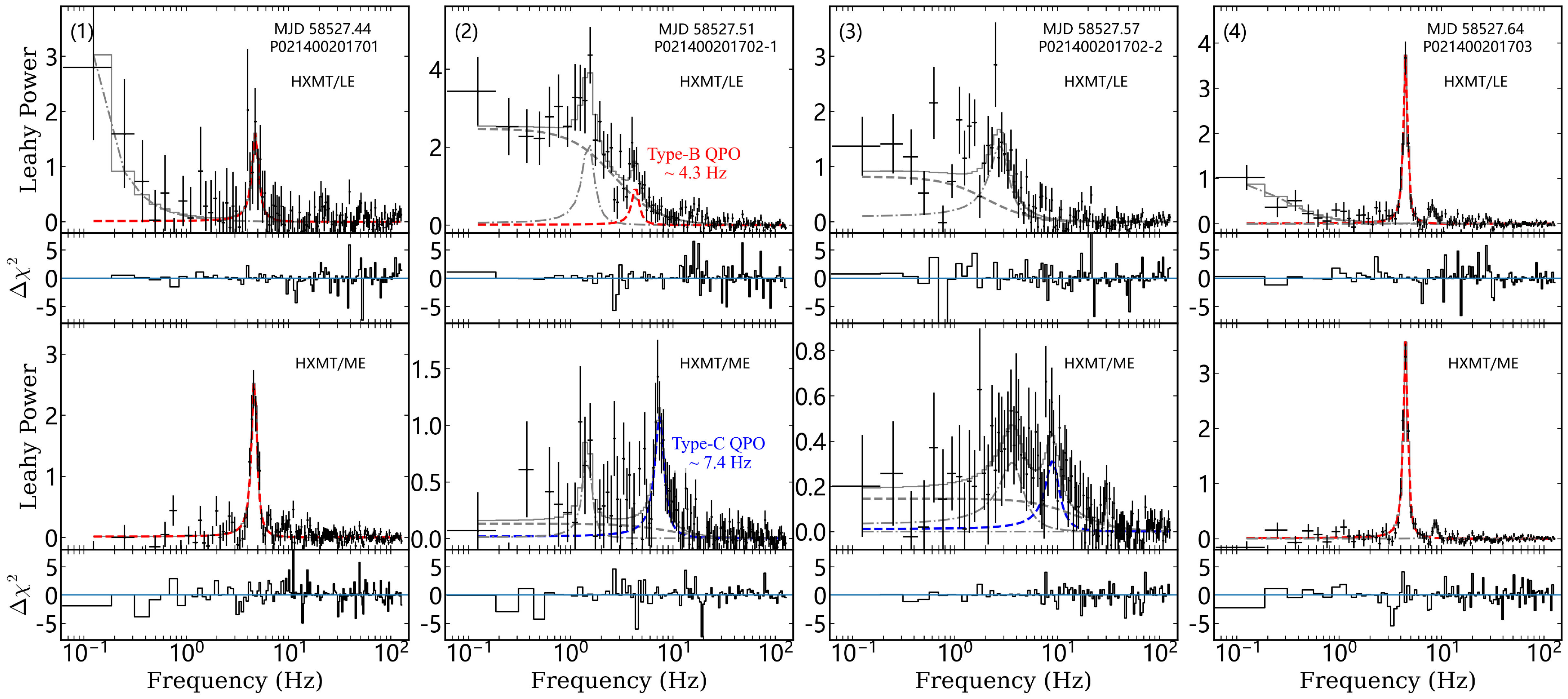}
\caption{Evolution of PDSs across the LE (upper) and ME (lower) bands from MJD 58527.4 to 58527.6. Type-B QPOs are marked with red dashed lines, and type-C QPOs are marked with blue dashed lines. From panel (1) to panel (2), the dominant noise type transitions to BLN (indicated by gray dashed lines), accompanied by the emergence of a $\sim$7.4 Hz type-C QPO (more prominent in ME) and the simultaneous detection of a $\sim$4.3 Hz type-B QPO (more prominent in LE).}
\label{fig:1702pds}
\end{figure*}

We used the pipeline of the \textit{Insight}-HXMT Data Analysis Software package (\texttt{HXMTDAS}) V2.05 to process the initial \textit{Insight}-HXMT data. After calibrating and screening the raw event data, we utilized tools within \texttt{HXMTDAS} (e.g., \texttt{helcgen, hebkgmap}) to generate source light curves and background light curves with a time resolution of 1/256 seconds. The net light curves are then calculated by using \texttt{lcmath}. For the \textit{NICER} data, we used the \texttt{HEASOFT} version 6.26 and \texttt{NICERDAS} pipeline to process the initial data, and produced light curves in 1--10 keV with a time resolution of 1/256 seconds using \texttt{XSELECT}.
We used the \texttt{powspec} tool with a segment duration of 64 seconds to generate average PDS.
Since the frequencies of QPOs during the HIMS and SIMS are relatively high ($\sim$ 1--11 Hz for type-C QPO and $\sim$ 4--5 Hz for type-B QPO), we employed an 8-second segment duration in place of 64-second segments to increase the signal-to-noise ratio (S/N) (more segments for averaging would reduce the uncertainty in PDSs).

All PDSs were normalized following the Leahy method \citep{Leahy1983}, and the Poisson noise was subtracted. To enable fitting with models in \texttt{XSPEC} (version 12.10.1), we processed the PDSs into a compatible format using the \texttt{flx2xsp} tool \citep{Ingram12}. We fitted the PDSs in \texttt{XSPEC} using a multiple \texttt{lorentz} model, with the best-fitting parameters and their 1$\sigma$ uncertainties determined from the chains produced by the Monte Carlo Markov Chain (MCMC) algorithm \texttt{emcee} \footnote{\url{https://github.com/zoghbi-a/xspec_emcee}}. The significance of the QPOs was quantified by the ratio of the integrated power under the \texttt{lorentz} component (i.e., the \texttt{norm} parameter) to its negative $1\sigma$ uncertainty (e.g., \cite{Motta_2015,zhangliang_2020,wangjingyi_2024,Kumar_2024}). Their fractional rms amplitude was calculated as \(\sqrt{\texttt{norm} / \langle C \rangle}\), where \(\langle C \rangle\) is the mean count rate of the light curve \citep{vanderKlis1988,VanDerKlis89}.

\section{Coexistence and interplay between type-C and type-B QPOs}
\label{sec:results}
As shown in \autoref{fig:QPO-evolution}, we present an overview of the outburst evolution of MAXI J1348-630. Before MJD (Modified Julian Date) 58517, the source was in the HS; it then transitioned to the HIMS, which lasted from MJD 58517 to 58522. From MJD 58522 to MJD 58540, the source was in the SIMS. After approximately 80 days in the SIMS and SS, the source entered the decay phase on MJD 58600. This evolution of spectral states in MAXI J1348-630 during its 2019 outburst has been the subject of detailed reporting and analysis \citep{Zhang2020,Alabarta_2022,You_2024}.

During the SIMS, type-B QPOs were extensively detected with a stable frequency of $\sim$4.5 Hz \citep[see \autoref{fig:QPO-evolution} and also][]{Belloni2020,Zhang2021,Liu_2022}. Notably, type-C QPOs were also detected several times during this phase with \textit{Insight}-HXMT \citep{Liu_2022}. These rare detections provide an important opportunity to explore the relationship between type-C and type-B QPOs. While previous studies suggested a rapid, mutually exclusive transition between type-B and type-C QPOs, our multi-energy-band analysis reveals that the two QPO types can briefly coexist in this source, as shown in the insets of \autoref{fig:QPO-evolution}.  In the following \autoref{sec:coexistence} and \autoref{sec:coexistence-combined}, we will present the detailed analysis.

\begin{figure}
\centering
\includegraphics[width=0.9\linewidth]{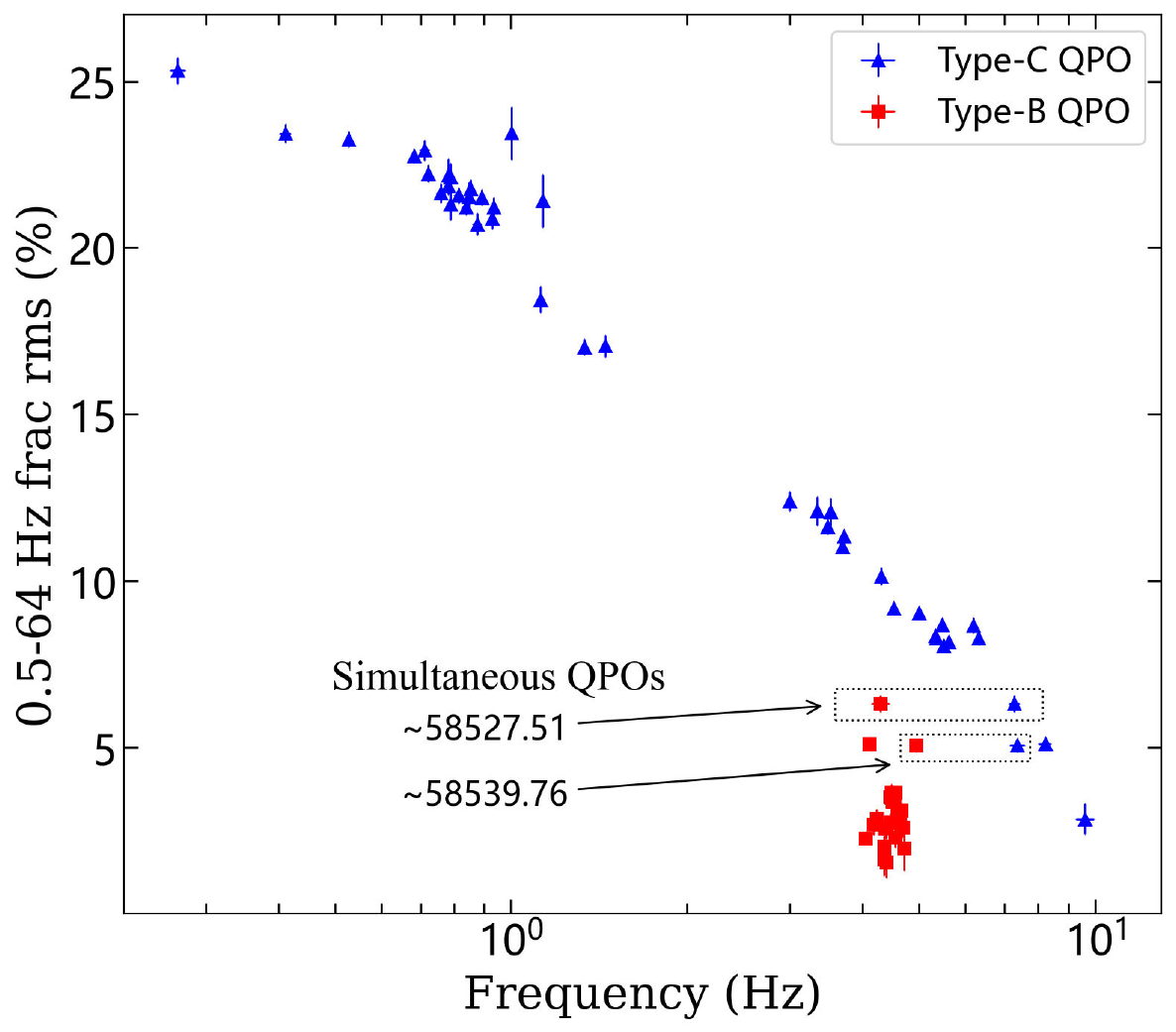}
\caption{Correlation between the frequencies of type-C and type-B QPOs and the total fractional rms during the HS, HIMS and SIMS of the outburst. Simultaneously detected QPOs are marked with black boxes.}
\label{fig:fre-rms}
\end{figure}

\begin{table*}[htbp]
  \centering
  \caption{Parameters of the type-B and type-C QPOs during the two epochs of simultaneous detections at MJD 58527.51 and 58539.76, respectively.} 
    \begin{tabular}{llllll}
    \toprule
    MJD   & Detector \&  Band & QPO Type & Frequency (Hz) & FWHM (Hz) & rms (\%) \\
    \midrule
    58527.51 & LE (1-10 keV) & Type-B & $4.29^{+0.14}_{-0.15}$ & $1.41^{+0.64}_{-0.49}$ & $2.67^{+0.45}_{-0.42}$ \\
\cmidrule{2-6}          & ME (10-30 keV) & Type-C & $7.44^{+0.23}_{-0.17}$ & $1.92^{+0.65}_{-0.49}$ & $10.48^{+1.25}_{-1.08}$ \\
    \midrule
    58539.76 & NICER (1-10 keV) & Type-B & $4.85^{+0.07}_{-0.08}$ & $0.33^{+0.27}_{-0.17}$ & $0.56^{+0.13}_{-0.10}$ \\
\cmidrule{2-6}          & ME (10-30 keV) & Type-B & $4.94^{+0.08}_{-0.08}$ & $0.47^{+0.40}_{-0.26}$ & $4.04^{+1.06}_{-0.77}$ \\
\cmidrule{2-6}          & ME (10-30 keV) & Type-C & $7.36^{+0.23}_{-0.22}$ & $1.32^{+0.11}_{-0.10}$ & $5.70^{+0.74}_{-0.88}$ \\
    \bottomrule
    \bottomrule
    \end{tabular}%
  \label{tab:Co-QPO-parameter}%
\end{table*}%

\subsection{Coexistence of type-B and type-C QPOs around MJD 58527.5}
\label{sec:coexistence}
When MAXI J1348$-$630 entered the SIMS at $\sim$MJD 58522, it exhibited typical timing features of this phase: the transition of the dominant noise type from BLN to power-law noise (PLN), the appearance of a type-B QPO, and the disappearance of the type-C QPO (see panel (b) of \autoref{fig:QPO-evolution}).
Approximately 5 days later ($\sim$MJD 58527.5), the dominant noise type returned to BLN, accompanied by the reappearance of a type-C QPO (see panel (2) of \autoref{fig:1702pds}).
As for the type-B QPO, it nearly disappeared in the ME band, leading to a mutually exclusive transition with the type-C QPO. However, the analysis of the LE band data from the same exposure reveals that the type-B QPO was likely still detectable and coexisted with the type-C QPO.

Specifically, in the exposure P021400201702, the LE band had only two short good-time-intervals (GTIs), lasting 300 seconds and 323 seconds, respectively. In the PDS of the first 300-second GTI, we detected a $4.29^{+0.14}_{-0.15}$ Hz QPO with a $3.2\sigma$ significance in the LE band (upper panel (2) of \autoref{fig:1702pds}), and during the same interval of ME band, a $7.44^{+0.23}_{-0.17}$ Hz QPO with a $4.9\sigma$ significance was detected (lower panel (2) of \autoref{fig:1702pds}). The fitting parameters for these simultaneous QPO detections are listed in \autoref{tab:Co-QPO-parameter}.  In addition, in both panels (2) and (3) of \autoref{fig:1702pds}, a broad component with a centroid frequency spanning approximately 1.5--3.5 Hz is clearly present. This component is observed multiple times during the SIMS when the noise is dominated by BLN (also see in the next \autoref{sec:coexistence-combined}). 

Clearly, the frequency of this 7.4 Hz QPO is non-harmonic with that of the 4.3 Hz QPO. The 4.3 Hz QPO is likely a type-B QPO according to its frequency, and the 7.4 Hz QPO is identified as a type-C QPO in \citet{Liu_2022}. To further clarify their classification, we analyzed the QPOs in the frequency versus total fractional rms (0.5--64 Hz) diagram—a method used for distinguishing QPO types \citep{Casella05,Alabarta_2022} and previously applied by \citet{Motta_2012} to report the first simultaneous detection of type-B and type-C QPOs in GRO J1655-40. As shown in \autoref{fig:fre-rms}, the analysis clearly identifies the 7.4 Hz QPO as type-C and the 4.3 Hz QPO as type-B. We also applied this method to distinguish type-C and type-B QPOs in another simultaneous detection, as described in the following \autoref{sec:coexistence-combined}.

In the next panel (3), the BLN remained the dominant noise type and the type-C QPO persisted, while the type-B QPO had disappeared. Approximately 6000 seconds later in panel (4), a strong type-B QPO returned, accompanied by the disappearance of both the BLN and the type-C QPO. This sequence of panels (1) through (4) demonstrates a coupled transition of the QPO types and the dominant noise. This transition pattern was observed again in the subsequent epoch (see \autoref{sec:coexistence-combined}).

\begin{figure*}
\centering
\includegraphics[width=0.9\linewidth]{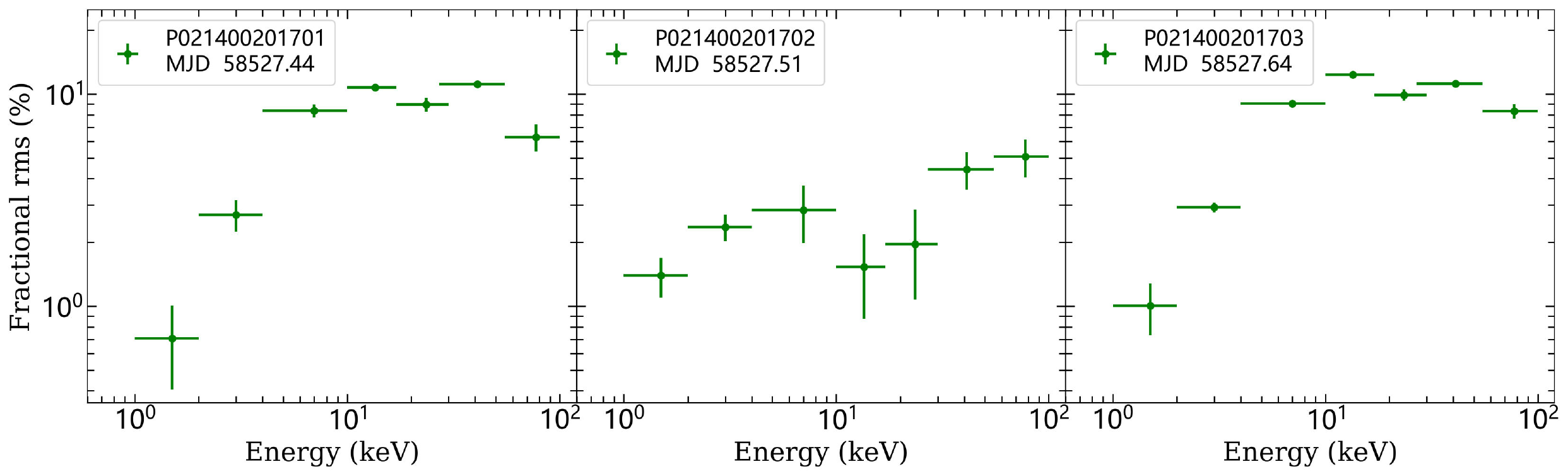}
\caption{Fractional rms spectra of type-B QPOs from three consecutive \textit{Insight}-HXMT exposures. Among these panels, the middle one corresponds to the observation where a type-C QPO was detected simultaneously; the other two panels represent the preceding and following exposures, in which only type-B QPOs were detected.}
\label{fig:rms-spectrum}
\end{figure*}

Additionally, as shown in panel (2) of \autoref{fig:1702pds}, the type-B QPO is more prominent in the LE band than in the ME band, unlike its usual behavior (see panels (1) and (4)). 
In order to quantitatively demonstrate the effect of the emergence of the type-C QPO, we studied the fractional rms evolution of type-B QPOs using more specific energy bands. The energy bands were defined as follows: 1--2 keV, 2--4 keV, and 4--10 keV (LE); 10--17 keV and 17--30 keV (ME); 27--55 keV and 55--100 keV (HE). As shown in \autoref{fig:rms-spectrum}, the rms amplitude in the hard X-ray ($>$10 keV) for exposure P021400201702 (where the two QPOs coexist) is significantly lower than in adjacent exposures where the type-B QPO was detected alone, with a reduction factor ranging from 1.4 (55--100 keV) to 7.5 (10--17 keV). 
This confirms that the presence of the type-C QPO modulates the strength and energy dependence of the type-B QPO, specifically in the ME band (see central panel of \autoref{fig:rms-spectrum}).

\subsection{A second epoch of QPO coexistence and its evolution around MJD 58539.8}
\label{sec:coexistence-combined}
The above analysis around MJD 58527.5 reveals that type-C and type-B QPOs coexist over a short period. 
A similar phenomenon was again observed when the type-C QPO reappeared around MJD 58539.8 \citep[see inset in \autoref{fig:QPO-evolution} (b) and also][]{Liu_2022}.

\begin{figure*}
\centering
\includegraphics[width=1\linewidth]{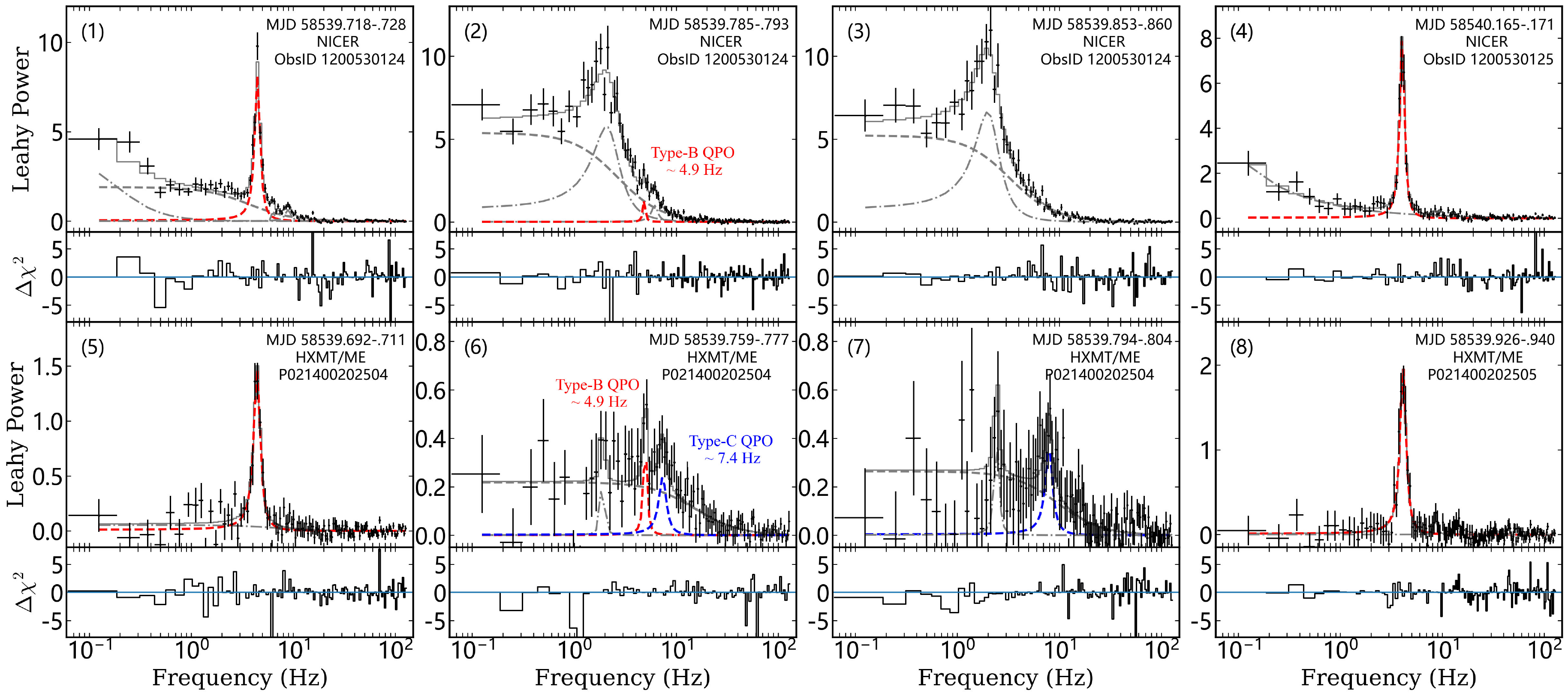}
\caption{Evolution of PDSs around MJD 58539.8 in the soft (\textit{NICER}, upper panels) and hard (ME, lower panels) X-ray bands. Type-B QPOs, type-C QPOs, and BLN are marked as in \autoref{fig:1702pds}. In second column (panels (2) and (6)), a $\sim$7.4 Hz type-C QPO appears, coinciding with a strong BLN, while a type-B QPO is simultaneously detected at $\sim$4.9 Hz.}
\label{fig:58539pds}
\end{figure*}

In \autoref{fig:58539pds}, we present the detailed evolution of PDSs around MJD 58539.8, covering the transition from a strong type-B to a prominent type-C QPO (see panels (5)--(7)).
Since \textit{NICER} also observed during this period, we used its soft X-ray data (1--10 keV) due to its higher count rate compared to LE. Panels (1)--(3) show \textit{NICER}'s PDSs from three consecutive GTIs of ObsID 1200530124, and panels (5)--(7) show the corresponding PDSs from the ME band (exposure ID P021400202504; observation duration $\sim$9600 s) at similar intervals.
Furthermore, to present the complete evolution sequence including the eventual returning of the type-B QPO, we have included data from subsequent observations: ME data from exposure P021400202505 (panel 4) and \textit{NICER} data from ObsID 1200530125 (panel 8).
Although the observation times of the two detectors are not perfectly aligned, their combined analysis enables tracking of the detailed evolution of the two QPO types during this period.

During the first GTI, a strong type-B QPO was detected in both \textit{NICER} and the ME band (see panels (1) and (5)), at similar frequencies of $4.46^{+0.02}_{-0.02}$ Hz ($16.7\sigma$) and $4.42^{+0.03}_{-0.04}$ Hz ($11.1\sigma$), respectively. For panel (1), the PDS additionally shows an unclassified QPO at $6.35^{+0.06}_{-0.07}$ Hz (persisted in panel (2) with a stable frequency of $6.15^{+0.11}_{-0.14}$ Hz) and the second harmonic of type-B QPO at $8.92^{+0.15}_{-0.12}$ Hz.
Additionally, the PDS in panel (1) exhibits a weak BLN component that peaks at $\sim$ 2 Hz, although BLN usually co-occurs with type-C QPOs. This appearance of the BLN component may be a sign of the transition of the PDS (as shown in subsequent panels).

During the second GTI ($\sim$1.6 hours after the first one), the PDSs underwent significant changes. For both \textit{NICER}’s and ME band’s PDSs (see panels (2) and (6)), the noise was dominated by BLN, and the type-B QPO was detected at $4.85^{+0.07}_{-0.08}$ Hz ($2.8\sigma$) and $4.94^{+0.08}_{-0.08}$ Hz ($2.6\sigma$), respectively. Notably, the significances of type-B QPOs were much weaker than those in the first GTI. An additional $\sim$2 Hz broad component was detected in panels (2) and (6), which has been mentioned in \autoref{sec:coexistence}.
Importantly, a QPO at $7.36^{+0.23}_{-0.22}$ Hz was simultaneously detected in the ME band with a significance of $3.2\sigma$. Its classification as a type-C QPO is confirmed by the \autoref{fig:fre-rms}. The evolution from the first to the second column in \autoref{fig:58539pds} shows that the emergence of the type-C QPO and the strengthening of the BLN coincided with a weakening of the type-B QPO. Moreover, the type-C QPO is not prominent in the soft X-ray ($<$10 keV), which is consistent with the result in the previous section (see \autoref{fig:1702pds}).

During the third GTI, the type-B QPO had vanished in both \textit{NICER} and ME bands. The type-C QPO became more prominent in the ME band ($3.8\sigma$), and its frequency increased to $7.88^{+0.23}_{-0.24}$ Hz.
Finally, in the last GTI of \autoref{fig:58539pds}, a strong type-B QPO reappeared, while both the type-C QPO and strong BLN vanished.

This detailed analysis of QPO evolution in \autoref{fig:58539pds} reveals a clear transition phase. The weakening and eventual disappearance of the type-B QPO occurred simultaneously with two key developments: the return of BLN as the dominant noise type and the emergence and strengthening of the type-C QPO. Conversely, the final reappearance of the type-B QPO was accompanied by the weakening of the BLN and the disappearance of the type-C QPO. This transition pattern is also observed in \autoref{fig:1702pds}.

\section{discussion}
\label{sec:discussion}
Type-C QPOs typically appear in the HS and HIMS and then disappear after the transition to SIMS, while type-B QPOs are characteristic of the SIMS. Our results demonstrate that type-C and type-B QPOs can briefly coexist during the SIMS, exhibiting significant interplay. This finding points to a previously unconsidered complex relationship between the two QPO types. By integrating our findings with prior research, we will discuss and propose a novel framework to describe this relationship. Subsequently, we will explore its implications for understanding the accretion flow evolution and its connection to jet ejection phenomena.

\subsection{Rethinking the relationship between type-C and type-B QPOs}
\label{sec:relationship}
Previous studies have described the relationship between type-C and type-B QPOs differently, likely due to variations in data analysis.
Some studies observed two QPO types separated by a very short interval, and thus described their relationship as "rapid transition" \citep{Casella04,Homan_2020,Liu_2022}.  In contrast, other studies have discovered evidence for their coexistence. The first such simultaneous detection was in GRO J1655$-$40, where the type-B and type-C QPOs were found to coexist for $\sim$400 seconds \citep{Motta_2012,Rout_2023}. Subsequent simultaneous detections in other sources, including XTE J1550$-$564 \citep{Motta_2014}, GRS 1915+105 \citep{motta2024}, MAXI J1820+070 \citep{Li_2025}, and Swift J1727.8$-$1613 \citep{pei_2025}, have further supported this phenomenon. Our analysis of MAXI J1348-630 provides robust evidence for two distinct episodes of coexistence between type-C and type-B QPOs.

The next question is whether the two relationships---"rapid transition" and "coexistence"---are mutually consistent or in conflict with each other. For instance, \cite{Homan_2020} reported a rapid transition from type-C QPO to type-B QPO in MAXI J1820+070 during the HIMS-to-SIMS transition (Figure 2(g), dynamic power spectrum: type-C vanished, type-B emerged at $\sim$4.5 Hz after $\sim$2000 s). However, \cite{Li_2025} detected the coexistence of both QPO types in the averaged PDSs from a short interval preceding the transition, indicating type-B QPOs had already existed but were only significant in averaged PDSs.

These two works demonstrate that in at least some cases reported as a "rapid transition", a brief coexistence phase can be identified on short timescales. This finding is consistent with our own results for MAXI J1348-630 presented in \autoref{fig:1702pds} and \autoref{fig:58539pds}. Furthermore, our detailed analysis reveals a clear energy dependence: while the type-C QPO dominates in the ME band, the type-B QPO is more prominent in the LE band.
Consequently, it is plausible that earlier studies, potentially constrained by the limited energy band coverage of their instruments, may have missed these subtle coexistence signatures. Therefore, employing a broad energy band is essential for a comprehensive investigation of this phenomenon.

However, a critical issue remains: is this coexistence of type-C and type-B QPOs merely a temporal evolutionary coincidence, or do they involve physical interactions?
Our analysis of MAXI J1348-630 reveals that the occasional emergence of type-C QPOs during the SIMS suppresses the strength of type-B QPOs, particularly in the hard X-ray bands. This suppression implies a competitive or interacting relationship between the two QPO types rather than independent variability, and suggests that they originate from distinct but interacting physical components.

Furthermore, while our analysis has focused on the competitive relationship between the type-C and the type-B QPOs, this relationship can extend to the BLN. As shown in \autoref{fig:1702pds} and \autoref{fig:58539pds}, the evolution of the BLN is a more prominent feature of the PDS than that of the type-C QPO. In fact, the BLN appears earlier than the type-C QPO and coexists with the type-B QPO in panel (1) of \autoref{fig:58539pds}. It then strengthens while the type-B QPO weakens and eventually vanishes (panels (2) and (3) of \autoref{fig:58539pds}), and finally, it nearly disappears in panel (4) where the type-B QPO becomes strong again. This pattern suggests that the competition is not only between the type-C and the type-B QPOs but, more prominently, between the BLN and the type-B QPO. Given that the BLN and the type-C QPO are typically coupled, this indicates two distinct variability modes represented by the BLN/type-C QPO and type-B QPO, respectively. Therefore, the competitive relationships we observe may reflect a competition between these two variability modes, which likely originate from distinct physical mechanisms, though the exact nature of these mechanisms requires further investigation.

\subsection{Probing accretion dynamics through the interplay between type-C and type-B QPOs}
\label{sec:insights_origin}

Type-C QPOs are commonly detected in the HS and HIMS, where the X-ray spectrum is dominated by hard X-ray emission from the Comptonization process. These QPOs have been observationally linked to the Comptonization spectral component \citep[e.g.,][]{Stiele2013, Gao_2023}. In contrast, type-B QPOs typically appear during the SIMS as the spectrum softens, and their initial emergence is often coincident with transient jet ejections \citep[e.g.,][]{Fender2009, Miller-Jones2012, Russell2019ApJ, Homan_2020}. The amplitudes of both QPO types exhibit a clear inclination dependence: type-C QPOs are stronger in high-inclination (edge-on) systems, while type-B QPOs are stronger in low-inclination (face-on) systems \citep{Motta_2015}. This dependence supports the interpretation that they originate from distinct physical components, with type-C QPOs associated with a horizontally extended structure such as the inner accretion flow or disk corona, and type-B QPOs linked to a vertically extended component like a jet \citep{Fender2009, Motta_2015,vandenEijnden2017}. 

Recent studies indicate that the HIMS/SIMS transition, which is typically accompanied by transitions between type-C and type-B QPOs, involves significant evolution in accretion dynamics. In MAXI J1348-630, multiple studies have supported that the corona evolves from a horizontal configuration in the HS and HIMS to a vertical one in the SIMS \citep{Garc2021,Bellavita_2022,Liu_2022,Alabarta24,Davidson_2025}. This coronal evolution pattern has also been suggested for other BHXRBs during state transitions \citep{Choudhury_2025,Wang2022,Cao2022}. Jet activity also changes dramatically during this phase: as the system transitions from HIMS to SIMS, the compact jet quenches rapidly, followed by a transient jet ejection (MJD $58521.5^{+1.8}_{-3.0}$) that nearly coincides with the state transition ($\sim$MJD 58522.5) \citep{Carotenuto2021MNRAS,Carotenuto2022MNRAS}. Similar jet behavior has been reported in other systems as well \citep{Homan_2020,Davidson_2025}.

Therefore, the transition between type-C and type-B QPOs offers a critical diagnostic window into accretion dynamics, as understanding their dynamic interplay is crucial for interpreting changes in accretion flow and jets. As discussed in \autoref{sec:relationship}, the transition can complete on a very short timescale, implying a highly dynamic or unstable flow configuration during this period.
Consistent with the general picture outlined above, the vertical expansion of the corona and the transient jet ejection coincide with the emergence of type-B QPOs, reinforcing the interpretation that type-B QPOs arise from vertically structured regions such as the jet or the vertically extended corona \citep{Belloni2020,Zhang2021,Liu_2022}. Conversely, the competitive relation between two QPO types suggests that type-C QPOs originate in a distinct, more horizontal Comptonizing region of the flow \citep{Rout_2023,Choudhury_2025}. In this scenario, a horizontally extended corona corresponding to the Comptonization process enhances hard X-ray emission above 10 keV, thus strongly reducing the rms amplitude of type-B QPO at this energy band. 
Furthermore, the occasional reappearance of type-C QPOs and their competitive suppression of type-B QPOs imply a scenario in which the horizontal Compton region can temporarily reform and suppress the vertical component, although the physical mechanism enabling this reconfiguration remains unclear.

Interestingly, while simultaneous detections of type-B and type-C QPOs are rare in BHXRBs, an analogous phenomenon is more often observed in neutron star (NS) Z-sources, where normal branch oscillations (NBOs;analogous to type-B QPOs) and horizontal branch oscillations (HBOs;analogous to type-C QPOs) can coexist in the averaged PDSs \citep{Motta_2017,Motta_2019}. The NBO in NSs is also associated with jet formation \citep[e.g.,][]{Migliari2007}. Such coexistence in NS systems typically occurs in the softer segment of the normal branch (NB) and coincides with radio-core flares \citep{Motta_2019}. This coexistence phase generally lasts only a few thousand seconds before the HBO disappears, leaving behind an isolated NBO \citep{Motta_2019}. These similarities between NSs and BHXRBs suggest that the mechanisms driving QPO formation are largely independent of the nature of the compact object. Instead, they likely reflect a universal geometric or dynamical competition within accretion flows and/or jets during state transitions. The short-lived coexistence of two QPO types thus serves as a key diagnostic of accretion flow reconfiguration, offering a unified framework for understanding variability across compact accretors.

\section{Conclusion}
\label{sec:conclusion}
In this work, we report the simultaneous detection of type-C and type-B QPOs in MAXI J1348-630 and reveal a significant interplay between them. We find that the emergence of the type-C QPO competitively suppresses the strength of the type-B QPO, particularly in the hard X-ray band ($>$10 keV), indicating a direct interaction between their underlying physical mechanisms rather than their independent variability evolution. Their brief coexistence, accompanied by energy-dependent suppression, reflects a transient and competitive reconfiguration of coronal structures and/or corona–jet coupling.


\section*{acknowledgements}
\label{sec:ack}
This work made use of the publicly available data and software from the \textit{Insight}-HXMT mission. The \textit{Insight}-HXMT project is funded by the China National Space Administration (CNSA) and the Chinese Academy of Sciences (CAS).
X.L.W. \& R.Y.M. were supported by the National Key R\&D Program of China (Grant No. 2023YFA1607902);
F.G.X. \& R.Y.M. were supported by the National SKA Program of China (no. 2020SKA0110102);
J.F.W. was supported by the National Key R\&D Program of China (Grant no. 2023YFA1607904);
This work was supported in part by the Natural Science Foundation of China (NSFC, grants 12373049, 12361131579, 12373017, 12192220, 12192223, 12033004, 12221003, U2038108 and 12133008).

\section*{Data Availability}
The data underlying this article are available in the \textit{Insight}-HXMT and \textit{NICER} public archive.



\label{lastpage}
\end{document}